\documentstyle[preprint,pre,aps,eqsecnum]{revtex}

\begin{document}
\draft


\title{A study of the phase transition in the usual statistical model for
nuclear multifragmentation}

\author{S. Das Gupta$^1$, and A. Z. Mekjian$^{1,2}$}

\address{
$^1$Physics Department, McGill University,
3600 University St., Montr{\'e}al, Qu{\'e}bec \\ Canada H3A 2T8\\
$^2$Physics Department, Rutgers University, Piscataway, New Jersey 08854\\ }

\date{ \today } 

\maketitle

\begin{abstract}
We use a simplified model which is based on the same physics as inherent in 
most statistical models for nuclear multifragmentation.  The simplified model 
allows exact calculations for thermodynamic properties of systems of large 
number of particles.  This enables us to 
study a phase transition in the model.  A first order phase transition can be 
tracked down.  There are significant differences between this phase transition 
and some other well-known cases.
\end{abstract}


\pacs{25.70-z, 64.60.My}



\section{INTRODUCTION}

Imagine that in heavy ion collisions because of two-body collisions 
thermalisation is obtained.  The compound system expands, reaches a volume
larger than normal nuclear volume and finally dissociates.  This dissociation
will carry the signatures of a thermalised system at this expanded volume.
This simple picture has been used to describe data obtained in heavy ion
collisions many times, sometimes with additional inputs.  In the late seventies
it was first used to describe composite production in the Bevalac data \cite
{Mekjian77}.  At the Bevalac energy the numbers of composites were small and
one was content to calculate cross-sections of light mass nuclei \cite {DM81}.
As experimental studies of heavy-ion collisions at lower energy progressed 
such models could be
tested in much greater details.  This prompted many detailed and elaborate
calculations.  Today large codes exist which will calculate cross-sections
of composites.  We mention two review articles \cite {Bondorf95},
\cite {Gross97} which provide relevant details and additional
references.

Another issue that has attracted a great deal of attention is whether or not
heavy ion collisions provide a window of opportunity for studying phase
transitiions in nuclei.  Mean field calculations \cite {JMZ83}show that 
nuclear matter has Van der Waal gas type behavior and thus in the $p-\rho$ 
plane there would be regions of liquid-gas coexistence.  It was conjectured
that during disassembly the nucleus could be in the coexistence region \cite
{Curtin83}.  Much of the interest in studying heavy ion collisions rests
on the hope that one may discern signatures of liquid-gas phase transition
in the data.  The recently developed lattice gas model for nuclear disassembly
addresses both the issues of fragment distributions ( the major achievemnt
of standard statistical model alluded above ) and liquid-gas phase transition
in a unified picture \cite {Pan95}.

The purpose of the present work is to investigate the possibility of a phase 
transition in a simplified version of the
statistical model \cite {Bondorf95}.  Given that the model with its many
elaborate features is able to reproduce significant details of data it is still
an interesting question to ask: does this have much relevance to a phase
transition which is one of the issues considered to be of primary interest
in heavy ion collisions.
This was briefly addressed in a recent paper \cite {DPKG97}
but we deal with this issue here with more rigour and in far greater detail.
Using a technique developed in a recent paper \cite {CM95} we are able to get 
exact results for systems as large as 2000 particles or more.  Another 
interesting feature is that we will be able to study ``percolative properties"
of the standard statistical model.  By this we mean the ``second moment" and
the size of the largest cluster, quantities which are of relevance in a
percolation model and the study of which is often very revealing 
\cite {Campi88}.

\section{THE MODEL AND THE CALCULATIONAL TOOLS}

In the simplified model investigated here we can have monomers or composites
of $k$ nucleons.  The composites have ground state energy 
$-W_0k+\sigma (T)k^{2/3}$.  The first term is the volume energy, with $W_0=$
16 MeV.  The second term is the surface tension term taken here as in \cite
{Bondorf95} to be temperature dependent :
$\sigma(T)=\sigma_0(\frac{T_c^2-T^2}{T_c^2+T^2})^{5/4}$ with $\sigma_0=$
18 MeV and $T_c=18$MeV.  In most of the calculations reported in this paper 
$\sigma $ is
temperature dependent.  However we also have done calculations with $\sigma$ 
temperature independent at constant value 18 MeV
and although values of specific heat, "boiling temperature", etc. change, no
qualitative changes are found.  At a finite temperature the composite can
be in the ground or one of its excited states.  Excited states of composites 
are taken into account using the Fermi gas model.  The internal energy of a 
composite at
temperature $T$ is $-W_0k+\sigma (T)k^{2/3}+T^2k/\epsilon_0$ where the last 
part comes from the population of excited states of the composite.    The value
of $\epsilon_0$ in the above expression is taken to be 16 MeV.
The intrinsic partition function of a composite of $k$ nucleons at temperature
$T$ is 
\begin{eqnarray}
z_k=exp[(W_0k-\sigma k^{2/3}+T^2k/\epsilon_0)/T]
\end{eqnarray}
It is assumed that the freeze-out density is small so that at disasembly 
different composites do not interact with each other (except through
excluded volume effect).

In the grand canonical ensemble, the average number of composite of $k$ 
nucleons is given by
\begin{eqnarray}
<n_k>=e^{\beta\mu k}\omega_k
\end{eqnarray}
with
\begin{eqnarray}
\omega_k=\frac{V_f}{h^3}(2\pi mT)^{3/2}k^{3/2}\times z_k
\end{eqnarray}
where for $k>1, z_k$ is given by eq.(1) and $z_1=1$.  The chemical potential
$\mu$ is fixed from the condition $\sum_{k=1}^Akn_k=A$.

Although not crucial for the work done in this paper we want to comment on the
volume parameter $V_f$ which appears in eqn (2.2).
We distinguish here between the volume $V$ to which the compound system has
expanded and the volume $V_f$ which is the available volume for the composite
of $k$ nucleons to move around.  For thermodynamics it is $V_f$ rather than 
$V$ which is to be used.   The two volumes are
related to each other but are not identical.  We assume that composites
appear with their normal volume.  Thus the volume which a composite of $k$
nucleons occupies is $k/\rho_0$ where $\rho_0=.16 fm^{-3}$.  It then follows  
that if the total number of nucleons fragmenting is $A$, the excluded volume
$V_{ex}$ is $A/\rho_0$ and we take the available volume for thermalisation to be
$V_f=V-V_{ex}$.  We show most of the calculations for $\rho=A/V=.05$.  This 
means calculations are done with $\tilde \rho=N/V_{f}=\frac{\rho}{1-\rho/.16}
=.0727$.

We wrote down the average number of particles according to the grand
canonical ensemble.  We will however mostly use the canonical ensemble using a
technique developed in \cite {CM95}.  The canonical partition function $Z_A$
for $A$ particles is given by
\begin{eqnarray}
Z_A=\sum \prod_{k\geq 1}\frac{\omega_k^{n_k}}{n_k!}
\end{eqnarray}
where $n_k$ is the number of composites which has $k$ nucleons and the sum above
goes over all the partitions which satisfy $\sum kn_k=A$.
The probability that a given admissible partition $ P(\vec n)= P(n_1,n_2,n_3...
...)$ is obtained is given by
\begin{eqnarray}
P(\vec n)=\frac{1}{Z_A}\prod \frac{\omega_k^{n_k}}{n_k!}
\end{eqnarray}
The partition function $Z_A$ can be built by recursion relation.  Starting with
$Z_0=1$ one can build all higher ones using
\begin{eqnarray}
Z_p=\frac{1}{p}\sum_{k=1}^pk\omega_kZ_{p-k}
\end{eqnarray}
Thus one has $Z_1=\omega_1;Z_2=\frac{1}{2}\omega_1^2+\omega_2$ and
$Z_3=\frac{1}{6}\omega_1^3+\omega_2\omega_1+\omega_3$ and so on.  The average
value of composite number in the canonical ensemble is readily obtained (this
is to be compared with eq.(2.2) of the grand canonical ensemble).  We have 
$<n_k>=\sum P(\vec n)\times n_k
=\frac{1}{Z_A}\sum \prod \frac{\omega_i^{n_i}}{n_i!}n_k$ which leads to
$<n_k>=\omega_k\frac{Z_{A-k}}{Z_A}$.  These results apply to any choice of
$\omega_k$ but in this paper we will focus only on the choice of eq.(2.3).
We like to emphasize that here the calculations in the canonical ensemble 
are done without any
Monte-Carlo sampling to obtain various quantities of interest.

In the canonical ensemble it is straightforward to isolate the the largest
cluster for each ``event".  Calculations in which a ``tag" is kept on the
largest cluster proved to be of significance in percolation
model analysis of data.  It is possible for us to calculate $<k_{max}>$, the
average size of the largest cluster.  
This should be scaled by the size of the system thus we compute $<k_{max}>/A$.
This turns out to be a very interesting quantity even in the statistical model.
Campi defined reduced moments as moments in which the largest cluster is
excluded.  Thus the second moment is $M_2=\sum_kk^2n_k-k_{max}^2$.  To compute
expectation values in which the largest cluster is fixed we need to compute the
partition function of such ensembles.  Clearly this partition function is given
by all the terms in the partition function which have
$\omega_{k_{max}}$ as the highest $\omega_k$ in the term.  Consider
\begin{eqnarray}
\Delta Z_A(k_{max})\equiv Z_A(\omega_1,....,\omega_{k_{max}},0,0,...)
-Z_A(\omega_1,...,\omega_{k_{max-1}},0,0,....)
\end{eqnarray}
We see that $\Delta Z_A$ is the partition function for ensembles with fixed
maximum cluster size $k_{max}$, since the first term collects all terms with
$\omega_{k_{max}}$ or lower, and the second term eliminates those terms which
do not have an $\omega_{k_{max}}$.  From this we can determine $Pr(k_{max})$,
the probability of obtaining a distribution whose largest cluster 
is $k_{max}$ \cite {CM95}.
\begin{eqnarray}
Pr(k_{max})=\frac{\Delta Z_A(k_{max})}{Z_A(\omega_1,........\omega_A)}
\end{eqnarray}
It is now straightforward to compute $<k_{max}>$ or the reduced moments.

Before we leave this section we like to comment on two of the possible
shortcomings of the model.  One is the assumption that different composites at 
the time
of dissociation are on the average far enough that they do not interact with 
each other.  The only effect that can be easily taken into account is through
the excluded volume  effect but the simple correction that we have alluded to
above is really very approximate and certainly not rigorous.  Trying to take 
interactions between different composites would change all the equations
for the partition function.  We can not correct this by simply deciding to take
an arbitrarily large freeze-out volume so that interactions between different
composites are indeed negligible because the major part of the physics
will have been already decided and that may pertain to a smaller volume.  In 
some sense, the lattice gas model \cite {Pan95} does a better job of
interactions between different composites.  The other approximation
is in the assumption that the internal partition function is given by eq. (2.1). 
That equation uses expressions for excitation energy and entropy
per particle which is valid only in the infinite non-interacting nuclear
matter limit.  Thus there will be pre-factor before the exponential in eq.(2.1)
which may completely mask the $k^{3/2}$ factor in eq.(2.3).  Other
choices for the $k$ dependence have been considered\cite {Ccm95}.  

\section{THE FREE ENERGY, THE LARGEST CLUSTER AND THE SPECIFIC HEAT}

The free energy of a system of $A$ particles is given by $F=-TlnZ_A$.  The
canonical partition function is directly calculated by the recurrence method
of eq.(2.6) and thus the
free energy is readily available.  For a system
of 2800 and 1400 particles this is shown in Fig. 1  as a function of
temperature.  The figure suggests that
there is break in the first derivative in $F$ which occurs at about 7.3 MeV
for 2800 particles and at 7.15 MeV for 1400 particles.  This means there is a 
first order transition here.  There will be a sudden jump in entropy
accompanied by an infinite specific heat at the ``boiling temperature".  Of
course although 2800 is a large number it is not the thermodynamic limit thus
the specific  heat would be large but not infinite.  The behavior of the
specific heat as the system size grows is shown in Fig. 2.  Notice the peak
becomes progressively high and narrow as the system grows larger.  This huge 
jump in specific heat at the ``boiling point" is readily understood if we
consider the behavior of $<k_{max}>/A$ for various size systems as function of
temperature.  Starting from a value higher than 0.5 it quickly drops to much
lower value as the ``boiling point" is traversed.  This drop becomes 
progressively sharper as the system size grows.  Calculations with finite 
systems (fig. 3) suggest that for infinite sytems $<k_{max}>/A$ will drop
from about 0.5 to at least very close to zero at the boiling point.  It is the  
sudden disappearance of the large blob at the ``boiling point" that is 
responsible for the specific heat going to infinity.

\section{ABSENCE OF A POWER LAW, EFFECTIVE $\tau$, AND THE SECOND MOMENT}

In a percolation model a power law emerges near the critical point.  The yields
$Y(A)$ of the composites obey $Y(A)\propto A^{-\tau}$.  In the lattice gas
model there is a continuous line in $\rho-T$ plane where such a power law
emerges.  Although the power law is expected to emerge only in the neighbourhood
of percolation point an effective $\tau$ can always be deduced from the data
using \cite {Pratt95}
\begin{eqnarray}
\frac{\sum_{n1}^{n2}AY(A)}{\sum_{n1}^{n2}Y(A)}=\frac{\sum_{n1}^{n2}AA^{-\tau}}
{\sum_{n1}^{n2}A^{-\tau}}
\end{eqnarray}
We have taken $n1=5$ and $n2=30$ which is similar to values used in other
analyses.
In the model pursued here we do not find the emergence of a power law.  This
is not just because we are using a finite number of particles.  One sees a much
clearer emergence of a power law in the lattice gas model even with smaller
particle number than what is considered here.  An example of yields $Y(A)$
in the model is demonstrated in fig 4.  In spite of a lack of a power law
below, at or above the boiling temperature
deducing an effective $\tau$ using formula (4.1)is useful.  As the 
temperature changes the value of the effective $\tau$ will go through a minimum
as shown in fig.5.  The minimum value depends upon the freeze-out density
$\tilde \rho$ used for the calculation.  For the case shown in figs. 4 and 5
the value of the $\tau_{min}$ was 1.84 but a lower value is found if a larger
value of the freeze-out density is used; correspondingly a larger value is
obtained if a lower freeze-out density is used.  We also recall when a true
power law emerges at a critical point the value of $\tau$ is restricted to be
between 2 and 3 \cite {Stauffer}.  There 
is no such restriction here because there is no power law.

We now consider the second moment $M_2$.  This will go through a maximum as
the temperature changes.  The maximum  of $M_2$ and the minimum of
$\tau_{eff}$ are very close to each other and they both occur close to the
maximum of $C_v$ (fig. 5).  Since experimentally a measurement of $C_v$ is
rather hard (see however \cite {Poch}) it is easier to locate approximately the 
``boiling point" by the minimum of $\tau_{eff}$ or the maximum of $M_2$.  The
value of $M_2$ does not grow with particle number but saturates around 11.6.

\section{BOILING POINT FOR INFINITE MATTER}

We return to figs. 1 and 2.  While the figures clearly show evidence of the
occurence of a first-order transition and the ever increasing values of the
specific heat it is difficult from these figures to determine the value of the
boiling point in the infinite matter limit.  This limit is more easily obtained
from the grand canonical ensemble and using the knowledge gained from fig.3
that as one crosses the ``boiling point" a large blob of the size $\approx A/2$
suddenly disappears.  Equations (2.1), (2.2) and (2.3) are combined to give 
\begin{eqnarray}
<n_k>=\frac{V_f}{h^3}(2\pi mT)^{3/2}k^{3/2}e^{\beta(\mu k+W_0k+T^2k/\epsilon_0
-\sigma(T)k^{2/3})}
\end{eqnarray}
for $k>1$ and $<n_1>=e^{\beta \mu}\frac{V_f}{h^3}(2\pi mT)^{3/2}$.  In the
infinite limit the term proportional to $k$ in the exponent must disappear
at the ``boiling point".  Hence we should have
\begin{eqnarray}
1=\frac{1}{\tilde \rho h^3}(2\pi mT)^{3/2}[e^{\beta\mu}+\sum_2^{\infty}k^{5/2}
e^{-\beta\sigma(T)k^{2/3}}]
\end{eqnarray}
where $\mu=-W_0-T^2/\epsilon_0$.  This gives for $\tilde\rho$=.07273 the
boiling temperature is 8.187 MeV.  The boiling temperature increases with 
increasing freeze-out density in this model.

\section{SUMMARY AND DISCUSSION}
We have taken a basic statistical model for nuclear multifragmentation and have
searched and found a first-order phase transition in the model.  This phase
transition is quite distinct.  Here the $C_v$ goes to infinity at the 
``boiling point" and the second moment $M_2$ reaches a maximum which is finite
in the thermodynamic limit.  This is to be contrasted with first order 
transition in the lattice gas model.  In that model the second moment goes to
infinity but $C_v$ merely goes through a discontinuous jump.  In so far as
the lattice gas model (which can be mapped on to an Ising Model) is a more
standard description of liquid-gas phase transition, this particular model 
considered here is
quite distinct.  One hall mark of this distinctiveness comes from the sudden
disappearance of $<k_{max}>/A$ which jumps from about .5 to nearly, if not
exactly,
zero as the boiling temperature is traversed.  If we identify the large cluster
as a liquid then we have to say, looking at fig. 3, that we can have a mixed
phase in which half or more of the total number of particles are in a liquid 
phase 
and
 the rest in gaseous phase; or, we can have all the particles in a gas
phase but we can not , for example, have one-third of the particles in the
liquid phase and the rest in the gaseous phase.  One has to remind oneself that
interactions in the model are treated in a rather undemocratic fashion.
Interactions within particles in a composite are treated (witness the apperance
of binding energy, surface terms and excited states of the composite) but
interactions between composites are treated less stringently.  This would make
the model quite unrealistic for high values of freeze-out density but of course
the model is designed for low freeze-out densities.

We find it satisfying that the maximum of $M_2$ and the minimum of $\tau_{eff}$
appear at a temperature close to that of the maximum of $C_v$.  Since the
measurement of $C_v$ is hard this means one may still locate approximately
the ``boiling temperature" by indirect means.  This remains true in the
lattice gas model in spite of rather different character of the phase transition
there.  Thus this seems to be a characteristic shared by many models.

\section{ACKNOWLEDGMENT}
 
This work is supported in part by the Natural Sciences
and Engineering Research Council of Canada and by {\it le Fonds pour
la Formation de Chercheurs et l'Aide \`a la Recherche du Qu\'ebec\/} and by
the US Department of Energy, Grant no. DE FG02-96ER 40987.

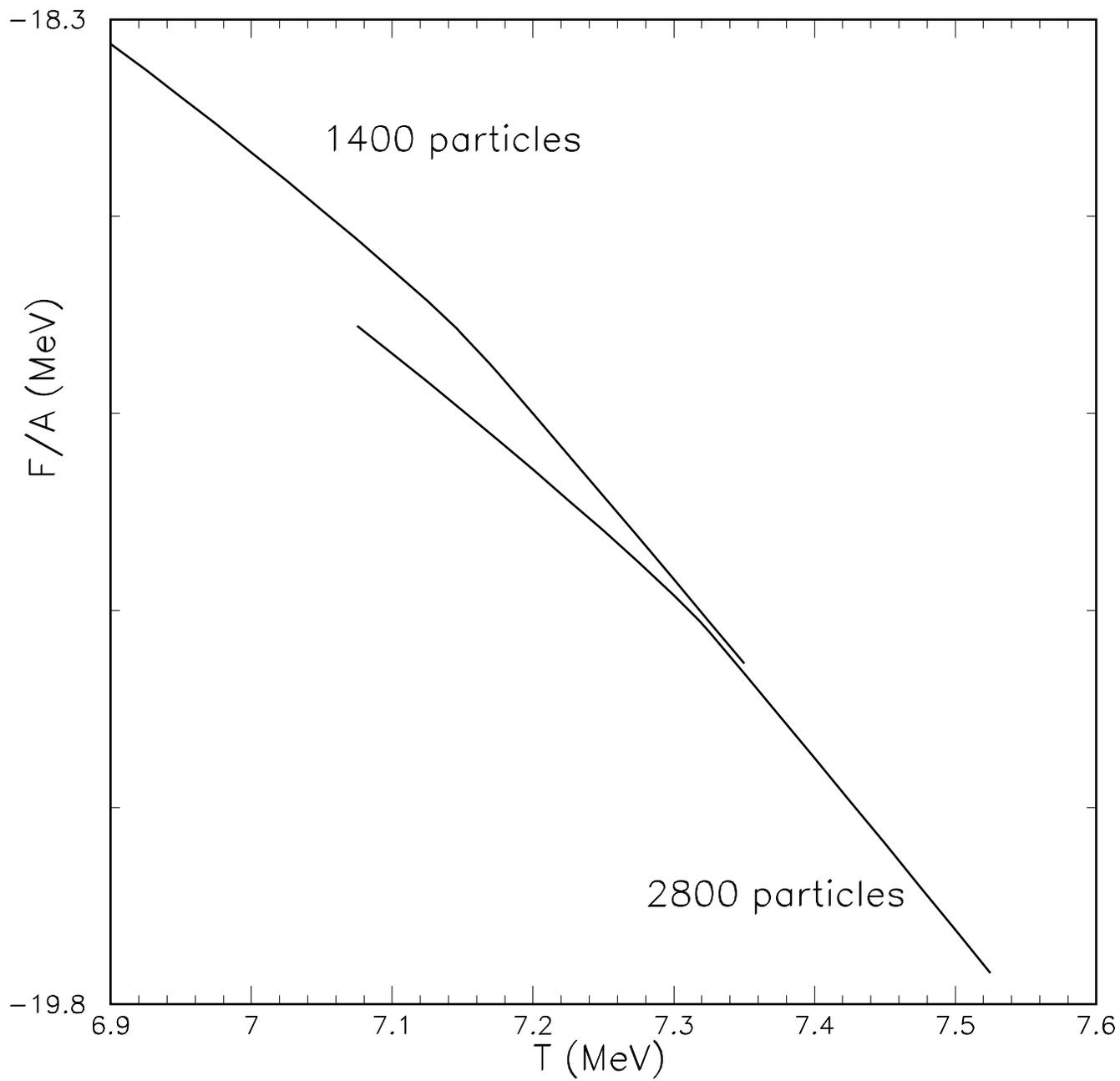
\begin{figure} \caption{ The free energy per particle for (a) a system of
2800 particles and (b) a system of 1400 particles plotted as a function 
of temperature.  The curves suggest that a break in the first derivative occurs
at 7.3 MeV and at 7.15 MeV respectively.} \label{phase_diagram} \end{figure}

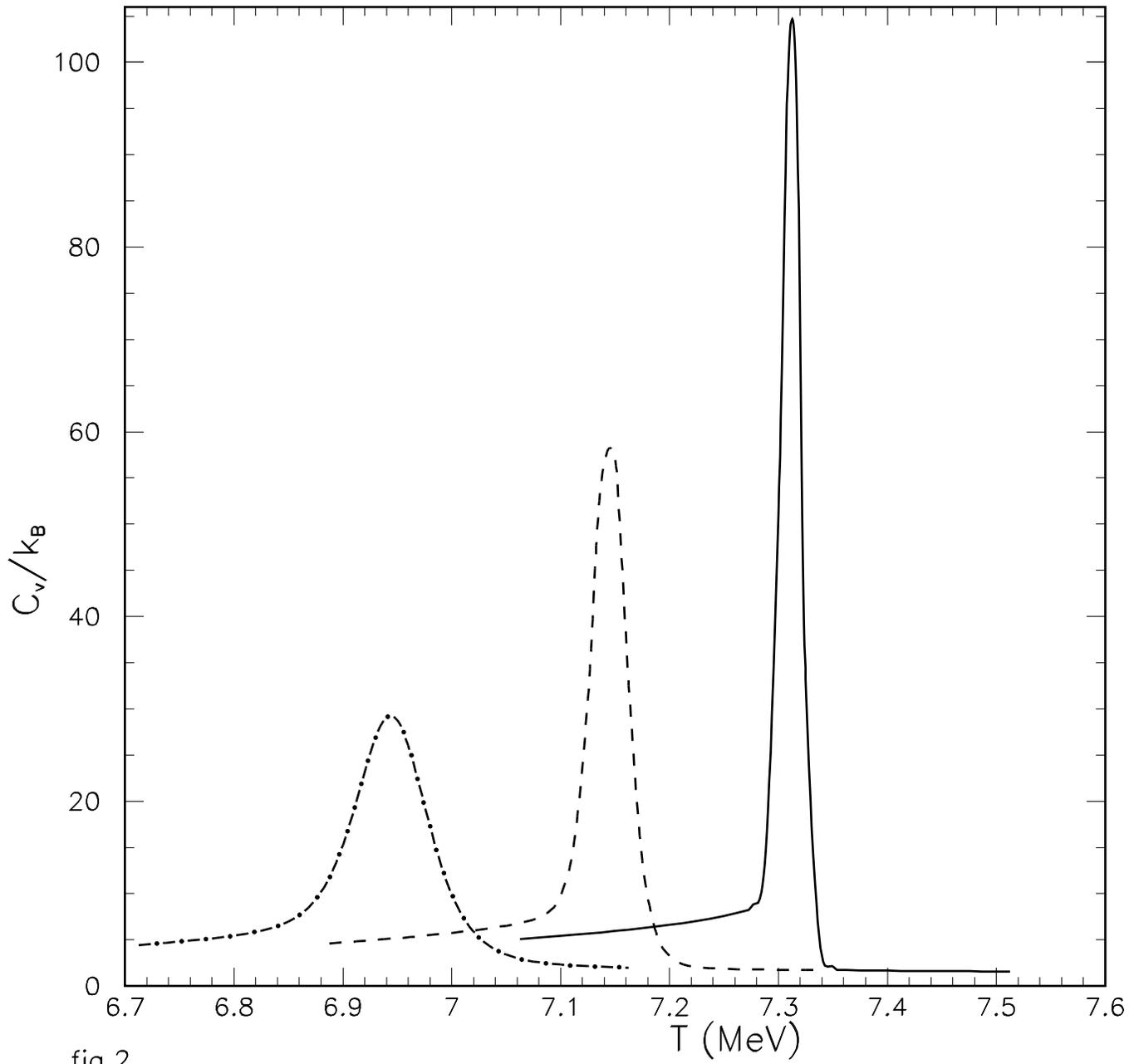
\begin{figure} \caption{ Curves for $C_v$ for 700 (dash-dot), 1400 (dash) and
2800 (solid) particles as a function of temperature.  The maxima occur at 6.95
Mev, 7.15 MeV and 7.3 MeV and coincide with what visually appear to be near
breaks in the first derivatives of free energy.  Notice that as the particle
numbers increase, the heights of the peaks increase and the widths narrow.  }
 \label{heat capacities} \end{figure}

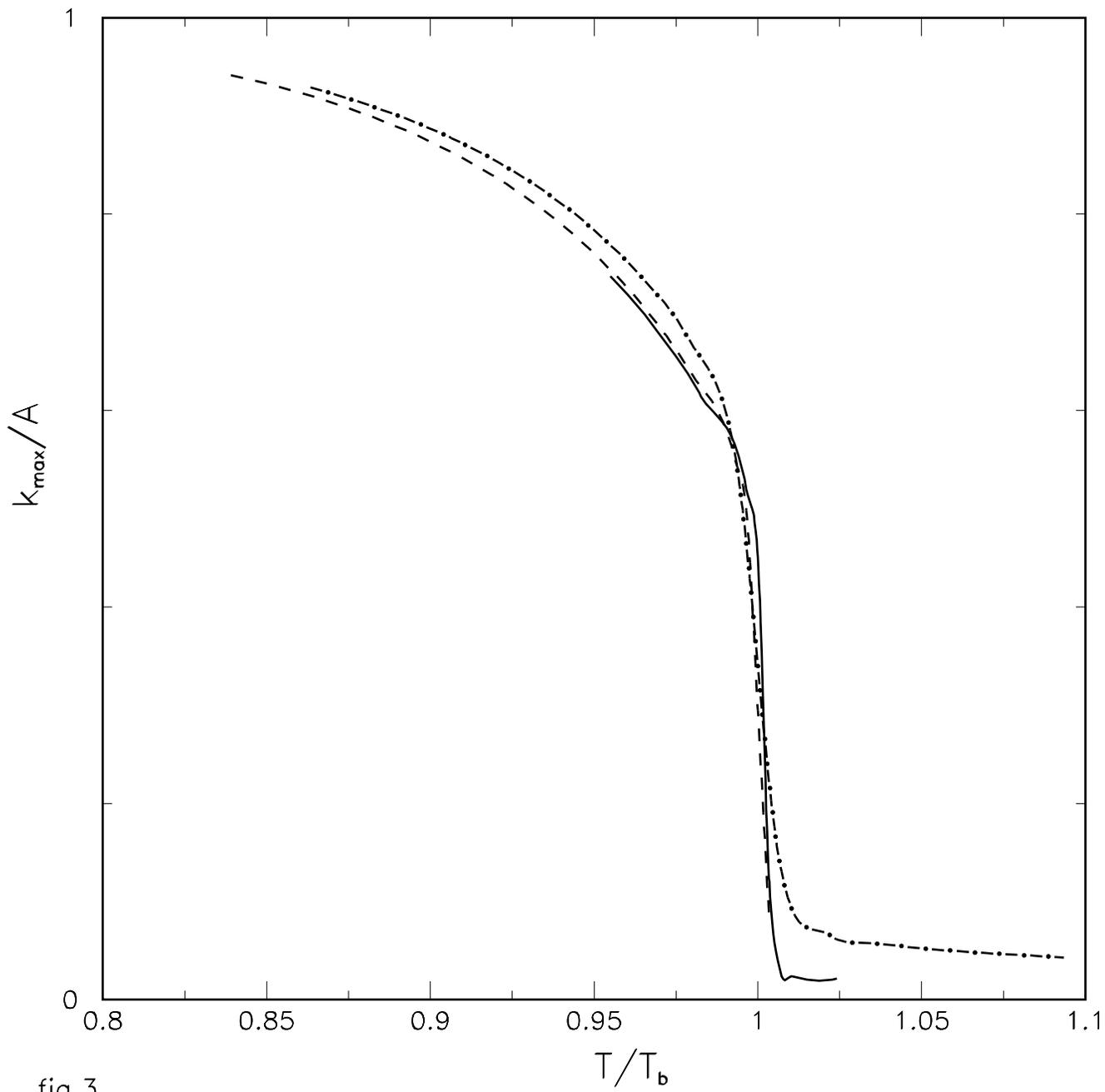
\begin{figure} \caption { Curves for $<k_{max}>/A$ as a function of the scaled
variable $T/T_b$ where boiling temparature $T_b$ is the temperature
at which $C_v$ is maximum.  The cases shown are for 700 particles (dash-dot),
1400 particles (dash) and 2000 particles (solid line).  The boiling temperatures
in the respective cases are 6.97 MeV, 7.15 MeV and 7.225 MeV.  Notice that
the steepness of fall increases as the particle number increases.  This sudden
fall begins at about 0.5 for $<k_{max}>/A$.  } \end{figure}

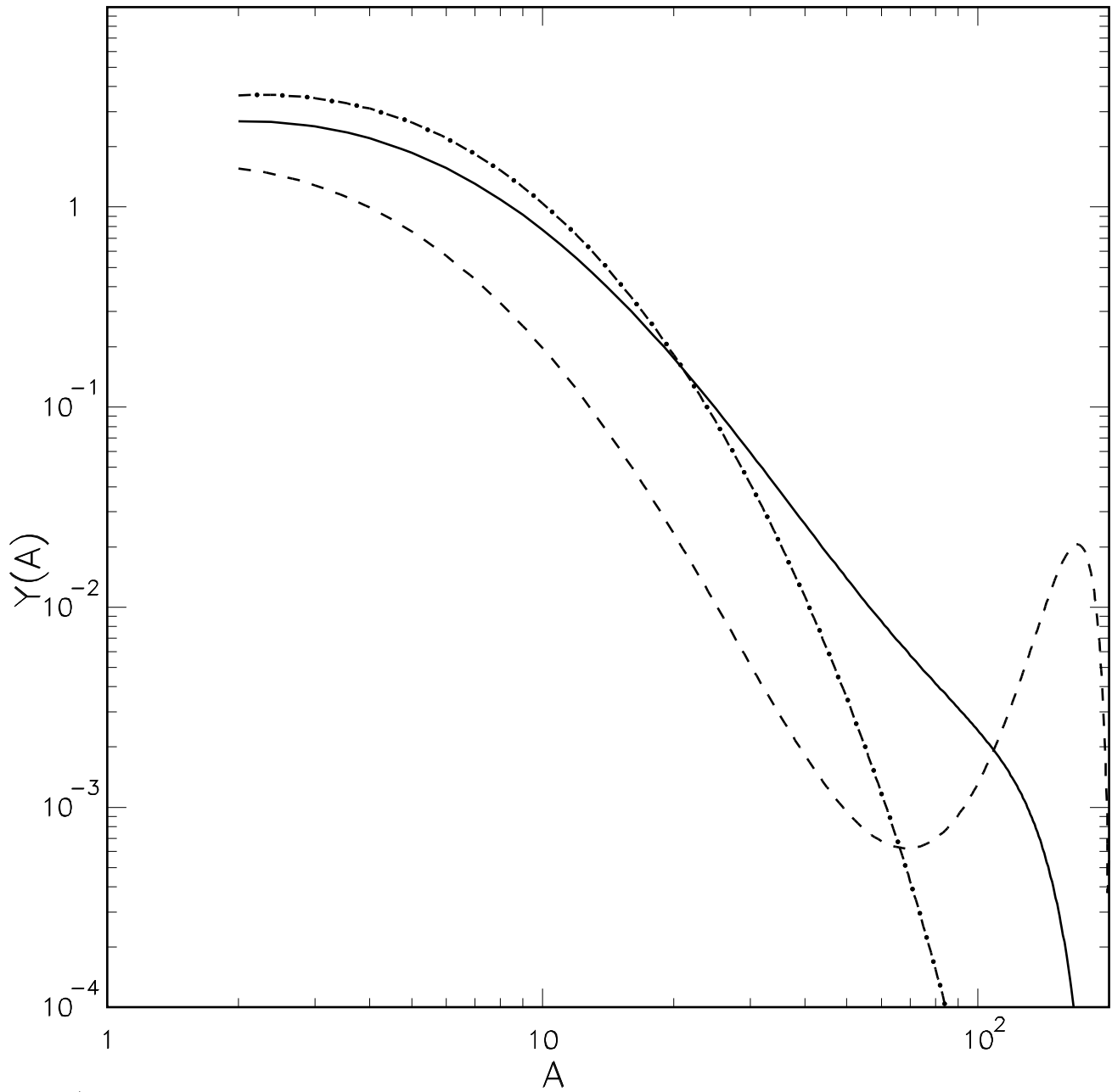
\begin{figure} \caption{ Curves for yields $Y(A)$ against $A$ at various 
temperatures.  The disassembling system chosen has 200 particles.  The
temparures chosen are 6.7 MeV (dash), 6.7 MeV (solid) (at this temperature
the effective $\tau$ is lowest) and 7.2 MeV (dash-dot). } \end{figure}

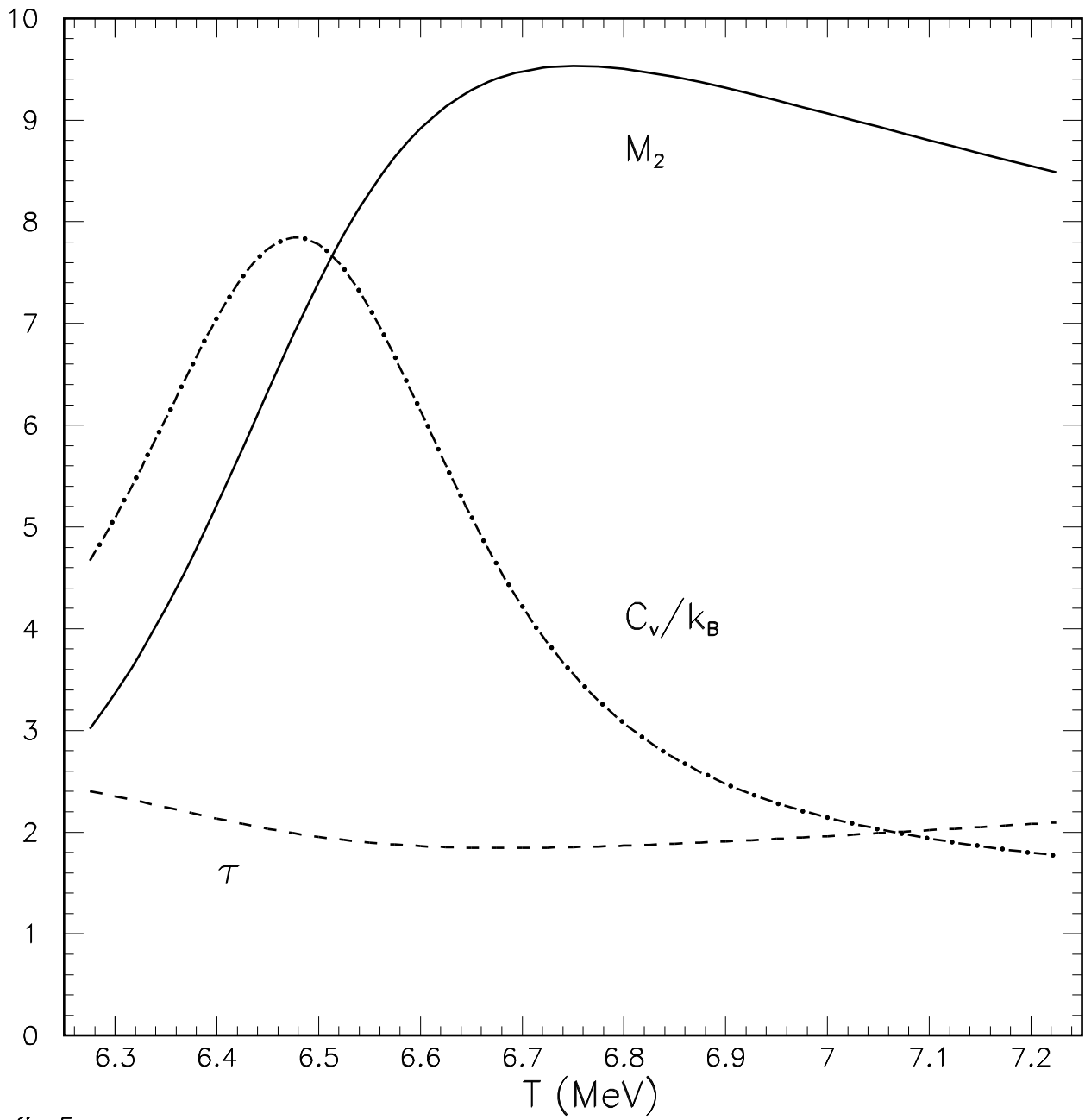
\begin{figure} \caption{ For a system of 200 particles curves for effective
$\tau$, specific heat and the second moment as a function of temperature. 
Notice that the minimum of $\tau$ and the maximum of $M_2$ are close to the
maximum of $C_v$.} \end{figure}

\end{document}